\documentclass[11pt]{article}

\usepackage[final]{acl}
\usepackage{booktabs} 
\usepackage{times}
\usepackage{latexsym}
\usepackage{svg}
\usepackage{tcolorbox}
\usepackage{booktabs,siunitx,threeparttable,makecell,colortbl}
\usepackage{xurl}
\usepackage{amsmath}

\newcommand{\sym}[1]{\textsuperscript{#1}}

\usepackage{amsmath,amssymb,mathtools}
\usepackage{algorithm}
\usepackage{algorithmic}
\usepackage{multirow}
\usepackage{makecell}
\usepackage{paralist}
\usepackage{array}
 \usepackage{subcaption}
 \usepackage[final]{microtype}

\newcommand{\grpo}{\textsc{GRPO}}

\newcolumntype{L}[1]{>{\raggedright\arraybackslash}p{#1}}
\newcolumntype{C}[1]{>{\centering\arraybackslash}p{#1}}
\newcolumntype{R}[1]{>{\raggedleft\arraybackslash}p{#1}}

\usepackage[T1]{fontenc}

\usepackage[utf8]{inputenc}

\usepackage{microtype}

\usepackage{inconsolata}

\usepackage{graphicx}

\newcommand{\acronym}{\textsc{QueStER}}
\title{{\acronym}: Query Specification for Generative Keyword-Based Retrieval}

\author{
  \textbf{Arthur Satouf}\textsuperscript{1,2,3,4}, 
  \textbf{Yuxuan Zong}\textsuperscript{4}, 
  \textbf{Habiboulaye Amadou-Boubacar}\textsuperscript{3}, \\
  \textbf{Pablo Piantanida}\textsuperscript{1,2}, 
  \textbf{Benjamin Piwowarski}\textsuperscript{4} 
  \\
  \textsuperscript{1} MILA – Quebec AI Institute \& ILLS, Canada \\
  \textsuperscript{2}Université Paris-Saclay \& CentraleSupélec \&  CNRS, France \\
  \textsuperscript{3}Air Liquide, France \\
  \textsuperscript{4}Sorbonne Université \& ISIR \& CNRS, France \\
  \\
   \small{
  \texttt{arthur.satouf(at)gmail.com, zong@isir.upmc.fr, habiboulaye.amadou-boubacar@airliquide.com,} }\\
  \small{
  \texttt{pablo.piantanida@cnrs.fr,  benjamin.piwowarski@cnrs.fr}
  }
}

\begin{document}
\maketitle
\begin{abstract}

Generative retrieval (GR) differs from the traditional index–then–retrieve pipeline by storing relevance in model parameters and generating retrieval cues directly from the query, but it can be brittle out of domain and expensive to scale. We introduce \acronym{} (\textbf{QUE}ry \textbf{S}pecifica\textbf{T}ion g\textbf{E}nerative Keyword-Based \textbf{R}etrieval), which bridges GR and query reformulation by learning to generate explicit keyword-based search specifications. Given a user query, a lightweight LLM produces a keyword query that is executed by a standard retriever (BM25), combining the generalization benefits of generative query rewriting with the efficiency and scalability of lexical indexing. We train the rewriting policy with reinforcement learning techniques. Across in- and out-of-domain evaluations, QueStER consistently improves over BM25 and is competitive with neural IR baselines, while maintaining strong efficiency.\footnote{
\href{https://github.com/arthur-75/quester}{Code on Github}.}


\end{abstract}

\section{Introduction}

Information Retrieval (IR) models aim to retrieve relevant documents from a large database that satisfy a user information need. 
They are a key component of search engines and of Retrieval-Augmented Generation (RAG) models~\citep{lewis2020retrieval}.
For a long time, IR BoW (bag-of-words) models, such as BM25~\cite{robertson1994okapi}, have been based on term-specific statistics,
allowing fast retrieval through the use of efficient index structures such as Block-Max WAND~\cite{grand_maxscore_2020}.
%
However, bag-of-words models suffer from the vocabulary mismatch problem, whereby the user query might contain keywords that do not appear in relevant documents. 

A standard way to solve this problem is to rely on query rewriting techniques. The first works~\cite{lavrenko2001relevance, abdul-jaleel2004umass} add keywords extracted from the  documents retrieved by the original query. This might lead to query drift, especially if the retrieved documents are not related to the user information need.
More recently, \cite{jagerman2023query, alaofi2023can} leveraged LLMs to improve the quality of the rewritten query. While improving compared to previous works, these strategies degrade when using a small LLM (e.g. smaller than 4B of parameters). When using large LLMs (with complex prompts), and moreover sampling several times~\citep{zhang2024exploring}, which decrease the efficiency. Exploring alternatives that balance effectiveness \emph{and} efficiency is important.

Nowadays, a common alternative to increase the effectiveness of IR models is to build models on top of Transformer~\citep{vaswani2017transformer} architectures.
Among the different models, dual encoders (dense and sparse), as well as late-interaction models such as ColBERT~\citep{santhanam2021colbertv2}, are the most effective and efficient first stage neural rankers. Other architectures such as cross-encoders~\citep{nogueira2019passage} are only used for re-ranking, and are not the focus of this paper. In addition, learned sparse lexical retrievers (e.g., DeepImpact, RRA)~\citep{basnet2024deeperimpactoptimizingsparselearned, satouf2025rationalretrievalactsleveraging} predict context-dependent term weights and sometimes lexical expansion, mitigating vocabulary mismatch while retaining inverted-index retrieval.

However, neural IR models have shortcomings since using a neural (first-stage) IR model efficiently requires indices (either dense or sparse, depending on the type of model). 
This is costly for two reasons. 
First, the space taken by those indices is much larger than that for BoW models (e.g. a dense index on MS MARCO with 8.8M documents using FP16 for 768-dimensional vectors requires 13 GiB, while the BM25 index only takes 0.67 GiB~\citep{park2025decoding}).
Second, when the model is re-trained, the index must be rebuilt, which is not practical with large collections such as that of the OpenSearch Euro project~\citep{granitzer2024impact}.

%

To tackle this issue, Generative Retrieval (GR) models attempt to internalize the index in the parameters~\citep{tay2022transformer}, which avoids entirely the construction of an index. During inference, these models directly predict a document identifier conditioned on the user query.
Some GR models rely on arbitrary document identifiers~\citep{tay2022transformer, wang2022neural}, but these models do not generalize well, especially for large collections.
Another research direction is to use document meta-data~\citep{zhou2022ultron, tang2023semantic, zhang2024generative} -- such as URL, title, or some keywords from the document -- as identifiers. While these approaches generalize better, they are still limited by the metadata they use.

In this paper, we generalize over meta-data based GR approaches, and posit that instead of trying to map queries to some document meta-data, \emph{generative models should generate search specifications}. A simple specification is a set of keywords that can be processed by e.g. a BM25 model, which resembles in this case query rewriting techniques. A more complex system would rely on structured specifications already present in major search libraries such as Lucene.\footnote{An example of a structured query specification can be found, see 
(\href{https://lucene.apache.org/core/2_9_4/queryparsersyntax.html}{Lucene docs}).}

The advantages are threefold. First, generating search specifications means that we can rely on established search technologies for which retrieval has been thoroughly optimized and for which query languages can be used to express complex user information needs. This leads to potentially more effective and efficient models than relying on document meta-data alone. Second, when the underlying neural network evolves, the index does not need to be rebuilt. Third, a user can analyze the query, which is a major concern in explainable AI-sensitive applications (e.g., law, patents, and medicine).






In this work, we propose a model named \acronym{} (for Search Specification Generative Retrieval). \acronym{} is based on an LLM whose parameters are optimized, with GRPO~\citep{shao2024deepseekmath}, to generate keyword queries for a search engine. In this work, we restrict the search engine to be a keyword-based BM25. Through our experiments, we show that \acronym{} is both effective (+4.0 nDCG@10 in-domain and +5.3 out-of-domain, compared to BM25) \emph{and} efficient ($\approx 28$ ms/query for BM25). \acronym{} reaches the performance of prompt-based methods that depend on much larger LLMs, despite using only a 4B backbone LLM. 


\section{Related Work}

\paragraph{Information Retrieval} 
Information Retrieval focuses on efficiently and effectively retrieving relevant documents that satisfy users information needs. Although current neural approaches such as DPR~\citep{karpukhin2020dense} or RepLLaMA~\citep{ma2023finetuning} dominate the leaderboards, traditional sparse models such as BM25~\citep{robertson1994okapi} are still widely considered by the community due to their excellent tradeoff between efficiency and effectiveness. 

\paragraph{Generative Retrieval} 
Models retrieve documents by directly generating their identifiers (DocIDs) using language models, bypassing the indexing stage. 
A first series of work ~\allowbreak\citep{tay2022transformer, wang2022neural, sun2023learning, yang2023auto, lee2023glen} proposes to associate with a document an arbitrary DocID which comes from a (hierarchical) clustering process or is learned. In both cases, these models have limited power of generalization and their effectiveness decreases with the number of documents in the collection.
%
An alternative is to use ``natural'' DocIDs, such as document titles~\citep{de2020autoregressive, chen2022gere, chen2022corpusbrain}, URLs~\citep{zhou2022ultron},
N-grams~\citep{bevilacqua2022autoregressive} or document summaries~\citep{tang2023semantic}. Although this offers good interpretability, meta-data are not a perfect representation of document semantic content, and the performance of such models still lags compared to other neural approaches on large-scale datasets~\citep{pradeep2023does}.

\paragraph{Query Rewriting}
Query rewriting~\citep{carpineto2012survey} is a common solution in IR to improve query precision and expressiveness. 
Statistical models such as BM25~\cite{robertson1994okapi} use
pseudo-relevance feedback methods~\citep{lavrenko2001relevance, abdul-jaleel2004umass} that rely on heuristics and/or statistical analysis of the initial query with its retrieved documents. These methods are often prone to query drift~\citep{mitra1998improving}, which limits their applicability.
%
Taking advantage of the performance of recent LLMs~\citep{yang2025qwen3, grattafiori2024llama, team2025gemma}, various LLM-based query rewriting methods have been proposed. Many works directly leverage the strong capabilities of LLMs to expand or rewrite queries through various prompting strategies, including natural language questions~\citep{alaofi2023can, ye2023enhancing}, keywords~\citep{jagerman2023query,li2024query,mackie2023generative} or even passages~\citep{gao2023precise,wang2023query2doc,shen2024large,zhang2024exploring}. 
While this leads to an increased retrieval effectiveness, successful models rely on complex prompts, very large models, and multiple samples, each being costly at inference time.
These drawbacks led to the development of approaches more closely related to ours, and which propose to finetune smaller LLMs.
Following~\citep{nogueira2017task}, several works~\citep{mao2024rafe, ma2023query, peng2024large, hsu2024grounding, yao2025llm} propose to leverage reinforcement learning techniques using a multi-source reward -- e.g. annotated samples, cross-encoder like MonoBERT~\citep{nogueira2019passage}.
%
However, these methods are used within a multi-hop scenario~\citep{hsu2024grounding}, with low efficiency, or are applied in a RAG system~\citep{mao2024rafe}, and did not really look the underlying IR performance (as well as using a weaker reward signal as in our work).


\section{Method}
\begin{figure*}[t]
    \centering
    \includegraphics[width=1\linewidth, trim=3.7cm 3.7cm 1.015cm 3.7cm, clip]{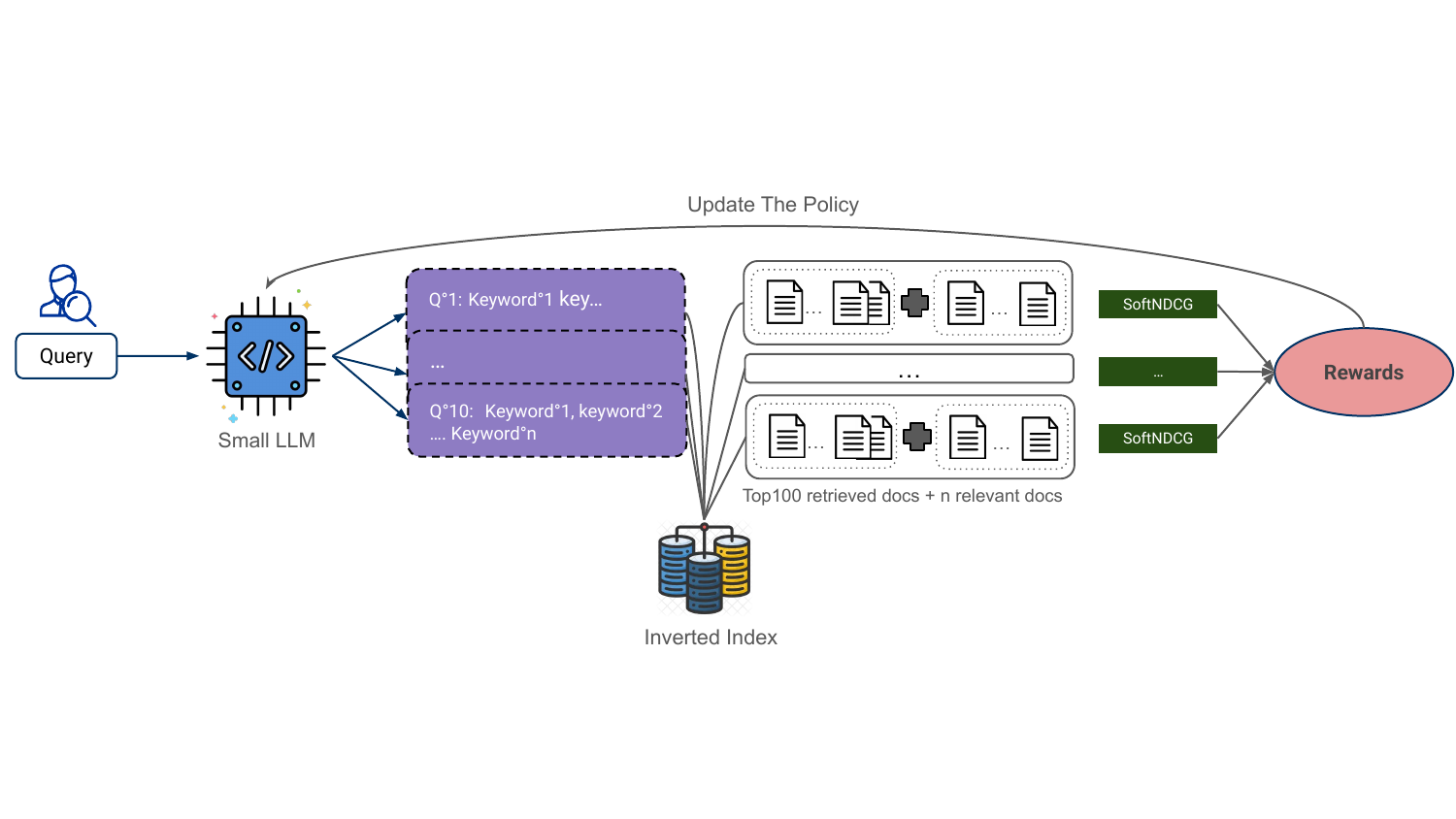}
    \caption{Overview of our query rewriting framework. A LLM generates multiple candidate queries, which retrieve documents using efficient index-based BoW IR models. The top-$k$ retrieved results are annotated with a cross-encoder reference from which the expectation of nDCG, SoftNDCG, can be computed. The resulting rewards are then used to update the policy for improved reformulation.}
    \label{fig:quester}
\end{figure*}
\newcommand{\query}{q}
\newcommand{\SE}{se}
\newcommand{\policy}{p_{\theta}}

\subsection{Problem Formulation}
Our goal is to build a model able to generate a query specification $\policy(\query)$ from an initial user query $\query$, where the re-written query leads to both efficient and effective retrieval.
%
%
To process the query specification $\policy(\query)$, we use existing BoW search engines $\SE$, e.g. BM25~\cite{robertson1994okapi}, to process the search specification $\policy(\query)$ depending on the search engine $\SE$. In this work, we use a bag-of-words model, namely BM25.

The search engine returns an ordered list of documents $\SE(\policy(\query))$, whose quality can only be evaluated using IR metrics, which prevents having a direct signal to learn $\policy$.
We thus frame the problem as a reinforcement learning problem, where the policy $\policy$ corresponds to the generation, and the reward corresponds to an IR metric.
More precisely, this metric is a scalar reward $R(\query, \policy(\query))$ that reflects the quality of the resulting query. Note that the reward should capture both the effectiveness and the efficiency of the resulting query.
The goal is to learn a policy that generates higher-reward candidates.

\subsection{Policy Optimization}
\label{sec:grpo}

We optimize the rewriting policy $\policy$ with Group-Relative Policy Optimization (\grpo)~\citep{shao2024deepseekmath}. 
For each query $\query$, the policy samples a group of $m$ candidates $\{q_i\}_{i=1}^m$, their associated rewards $\{r_i\}$ are calculated as described in Section~\ref{sec:reward}, and group-relative advantages $a_i = r_i - \bar r$ (with $\bar r=\tfrac{1}{m}\sum_i r_i$) drive a clipped policy-gradient update. 
Following prior large-scale GRPO practice~\citep{hu2025open}, we standardize rewards within each group and set the KL weight to $\beta=0$ (omit an explicit KL penalty) to encourage exploration (ablations in Section~\ref{sec:experiments} justify this choice).

\paragraph{Prompt.}
We conducted a prompt search (50+ candidates) to identify a minimal instruction that performs well zero-shot. Using a 1k-query subset of the MS MARCO training set, we selected the prompt that respects our keyword-writing format and maximized nDCG@10 with the Qwen3-4B policy. Our final prompt is:

\begin{tcolorbox}[
    colback=green!7,   
    colframe=black,    
    boxrule=0.5pt,     
    arc=2pt,           
    left=4pt, right=4pt, top=4pt, bottom=4pt
]
Generate relevant single-word keywords to improve retrieval performance. Only output unique keywords, separated by commas. [QUERY]: \texttt{\{query\}} [KEYWORDS]:
\end{tcolorbox}

\subsection{Reward}
\label{sec:reward}

To drive the RL algorithm, our reward should be sensitive to any change in the query, even if it marginally improves effectiveness or efficiency. Note that in this work, we did not include any reward component related to efficiency, as our first experiments did not show any difference with efficiency-based rewards.



\subsubsection{Soft and Hard nDCG}\label{softvshard}

To comply with those properties, we need a reward that is sensitive not only to the order between documents but also to the magnitude of the search engine scores $\SE${}. More precisely, if a relevant document is ranked after a non-relevant one, we still want to have a higher reward for the query where the difference in the scores is low rather than large.
%

An IR metric that matches our needs is SoftRank, which computes the expected nDCG~\cite{taylor2008softrank} as
$$
\mathbb{E}(nDCG@k) = \mathbb{E}\left(\sum_{i=1}^{k} \frac{\mathop{relevance}(d_i)}{\log_2(1 + K_i)}\right)
$$
where $\mathop{relevance}(d_i)$ is the relevance of the document $d_i$ to the query at hand, and $K_i$ a random variable corresponding to the rank of document $i$. SoftRank is based on the assumption that the scores returned by the search engines are normally distributed, i.e. $S(\query,d) \sim \mathcal{N}(\SE(q,d), \nu Id)$ where $\sigma$, the standard deviation, is a hyper-parameter that does not depend on $d$ or $\query$. 
%

Another way to see how this impacts the IR metric is to understand when a document $d_i$, with score $s_i$, is ranked before a document $d_j$ with score $s_j$:
\begin{equation}
p(d_i \succ d_j) \;=\; \sigma\!\left(\frac{s_i - s_j}{\nu}\right),
\end{equation}
where $\sigma$ is the sigmoid function and $\nu$ the standard deviation.  As $\nu \!\rightarrow\! 0$, the model tends to behave as nDCG as the rank is directly driven by whether $s_i>s_j$ (the above probability tends towards 0 or 1). When $\nu \!\rightarrow\! \infty$, $p(d_i \succ d_j) \rightarrow \frac{1}{2}$ so that no score impacts $\mathbb{E}(nDCG)$. In practice (see \S\ref{Variance}), a moderate $\nu$ yields the best trade-off between stability and discriminability. We name the two metrics HardNDCG (when $\nu\rightarrow 0$) and SoftNDCG (when $\nu$ has a moderate value).

An alternative metric would be to use the InfoNCE loss~\cite{oord2018representation}, commonly used to train neural IR models. However, in our preliminary experiments the training was unstable compared to the one using SoftRank~\citep{taylor2008softrank}. 

\subsubsection{Distillation: Cross-Encoder Assessments}

The dataset used most often to train IR models is MS MARCO~\citep{bajaj2016ms}, where the evaluations of whether a document is relevant to a given user query are based on user clicks. 
User clicks are known to be incomplete~\cite{Gupta_2022}: many queries are associated with a single relevant document, i.e. MS MARCO contains many \emph{false} negatives. In our case, $\mathbb{E}(nDCG)$ can be zero for all query candidates, which cannot be used for GRPO to learn properly.
To tackle this issue, we use distillation~\cite{hofstatter_efficiently_2021}, a standard technique used to train IR models that takes advantage of the most powerful IR models, namely cross-encoders (CE), to indicate whether or not a document is relevant.

\section{Experiments, Results and Analysis} \label{sec:experiments}
In this section, we first describe the experiments we developed to validate our model. Then, we analyze our experimental results, including the main result, the ablation studies, as well as some analysis of the efficiency.

\subsection{Experimental Setup}

\paragraph{Datasets}
We train our model on the MS MARCO v1 passage dataset~\citep{bajaj2016ms}. This dataset contains 8.8M passages. 
We evaluate the model \emph{in-domain} on the MS MARCO dev set (6980 queries), as well as the TREC DL 2019 (43 assessed topics) and TREC DL 2020 (54 assessed topics) sets. 
Following common practice, we report out-of-domain (OOD) performance of our method using the BEIR dataset~\citep{thakur2021beir}. 

\paragraph{Baselines}
We consider the following baselines for our experiments, including the neural dual models: 
\begin{compactitem} 
    \item ANCE~\citep{karpukhin2020dense}, a state-of-the-art dense neural IR model;
    \item SPLADEv2~\citep{formal2021splade}, a state-of-the-art sparse neural IR model;
    \item ColBERTv2~\citep{santhanam2021colbertv2}, a state-of-the-art multi-vector retrieval model, which is especially good at generalization. 
\end{compactitem}
We also compare with the BM25 variants, with or without query rewriting:
\begin{compactitem}
    \item A standard IR baseline, BM25~\cite{robertson1994okapi}, which also serve as the retrieving model for our rewritten queries. 
    \item RM3~\cite{lavrenko2001relevance}, a query rewriting model based on relevance feedback for BM25 retrieval.
    \item HyDE~\cite{gao2023precise}, a query rewriting model based on the generation of relevant passages;
    \item Query2Doc~\cite{wang2023query2doc} (abbr. Q2D), similar to HyDE, but that concatenates the query for BM25 retrieval. 
    \item LameR~\cite{shen2024large}, a LLM-based query rewriting model that concatenates the top retrieved documents. 
    \item MuGI~\cite{zhang2024exploring}, a state-of-the-art work that balances the query and pseudo-document for query rewriting. We report both the result based on ChatGPT3.5 and ChatGPT4.
\end{compactitem}
The results for SPLADEv2 and ColBERTv2 are copied from ~\citep{santhanam2021colbertv2}. The results for ANCE, RM3 and Q2D are copied from~\citep{wang2023query2doc}. The other baselines are from~\citep{zhang2024exploring}.
We report very few generative retrieval baseline results as they do not usually evaluate on BEIR and on the \emph{full} MS MARCO dataset.\footnote{We only report  MINDER~\citep{li2023multiview} for which a few results on full MS MARCO are reported.}
Among all the results reported in these papers, we generally select the ones reported in the tables by using the configuration put forward by the author. For the neural dual encoders, we take the result of the model trained \textbf{with} distillation~\citep{hofstatter_efficiently_2021}, as we are also distilling from a cross-encoder.

\paragraph{Implementation details}

We conduct all our experiments\footnote{Huggingface Transformers 4.53.3, TRL 0.19.1} with NVIDIA RTX A6000 48GB gpus (total cost \$150 - \$200). For GRPO training, we fine-tune Qwen3 models~\cite{yang2025qwen3} model\footnote{\href{https://huggingface.co/collections/Qwen/qwen3-67dd247413f0e2e4f653967f}{QWEN3}} across various size (0.6B, 1.7B and 4B) with LoRA~\citep{hu2022lora} (rank $r=40$, scaling $\alpha=40$, ``thinking'' mode disabled). 
The 4B version is our main model as it offers the best efficiency/effectiveness trade-off. 
For each input query, we sample from the LLM a group of $10$ candidates with a temperature $\tau=1.2$. The model is trained on $96{,}000$ randomly sampled queries from MS MARCO training set. As for the assessments, we either use the original assessments from MS MARCO train set, or the cross-encoder ones from OpenSearch \footnote{\href{https://huggingface.co/datasets/opensearch-project/msmarco-hard-negatives-llm-scores}{OpenSearch Project – MS MARCO Hard Negatives LLM Scores}}. 
Cross-Encoder-based relevance assessments are normalized by dividing the maximum score by the maximum score for a given query, leading to values in the range $[0,1]$.

For optimization, we use the AdamW optimizer with a learning rate of 5e-6, a batch size of 320 (using  20 micro-steps for one gradient update), and trained for one epoch. 
Our largest model (4B) takes around 2 days to train on a single GPU. 
When evaluating, we generate one rewritten query with greedy decoding (temperature $\tau=0$). During inference, processing the MS MARCO dev set (6980 queries) takes only around 5 minutes (batch size of 256).
The model parameters as well as the code will be made publicly available.

\subsection{Results and Analysis}

In this section, we perform a quantitative and qualitative analysis based on our experimental results.

\subsubsection{Main results}

In this section, we present the results of our model \acronym{} which is based on a Qwen-4B backbone, using cross-encoder assessments, a standard deviation $\nu=0.5$ when computing SoftNDCG and a cut-off at 10,000.\footnote{Note that because labeling with cross-encoder is costly, we assume that a document is not relevant if not in the top-100 re-labeled by the CE}

\paragraph{In-domain}

On MS MARCO Dev and TREC DL’19/’20 (Table~\ref{tab:indomain}), our GRPO-trained policy \textbf{\acronym{}} improves over BM25 and Qwen3-4B zero-shot prompting.  Compared to more closely related approaches (Q2D, HyDE, LameR, MuGI), \acronym{} has strong effectiveness while avoiding their generation overhead and instability. Compared to the state-of-the-art generative retrieval model like SEAL and MINDER, our model also achieve a better effectiveness, which illustrate our proposed methods could perform better to large scale dataset.  
Finally, compared to dual encoders (ANCE, SPLADEv2, ColBERTv2), which have been trained on MS MARCO, \acronym{} -- as the other query rewriting approaches -- have a much lower nDCG@10 (on DL’19 and ’20, is 71.7 for SPLADEv2 vs 62 for \acronym{}) but a competitive R@1K (78.9 vs 81.8).

\paragraph{Out-of-domain.}


On OOD datasets BEIR~\cite{thakur2021beir} , \acronym{} outperforms all the neural IR baselines (SPLADEv2, ColBERTv2) as reported in Table \ref{tab:statofart} -- showing the interest of such approaches for OOD retrieval. It is also competitive with query rewriting models (e.g., MuGI/GPT4 and Query2Doc), while being more efficient (see Section~\ref{sec:efficiency}). By contrast, prompt-heavy methods (HyDE, MuGI, and LameR) rely on large proprietary LLMs and stochastic decoding (temperature $\tau>0$), so their outputs are expensive to compute (several subsets are omitted due to the generation cost) and scores vary across runs. We instead decode with temperature $\tau=0$, producing deterministic, fully reproducible rewrites.

\begin{table}[!ht] 
\renewcommand\cellalign{l}
\renewcommand{\arraystretch}{1.1}
\centering
\scriptsize
\begin{tabular}
{m{1.3cm}C{0.55cm}C{0.55cm}C{0.55cm}C{0.55cm}C{0.55cm}C{0.55cm}}
\Xhline{1pt}
\multirow{2}{*}{model} & \multicolumn{2}{c}{\textbf{Dev set}} & \multicolumn{2}{c}{\textbf{DL19}} & \multicolumn{2}{c}{\textbf{DL20}}  \\
& \tiny RR@10 & \tiny R@1K &\tiny nDCG@10 & \tiny R@1K & \tiny nDCG@10 & \tiny R@1K \\
\Xhline{1pt}
\multicolumn{7}{l}{\emph{Dual encoder models}} \\
\Xhline{0.3pt}
ANCE & 33.0  &	95.9	 &64.5 &	75.5	 &64.6	 &77.6\\
SPLADEv2 & 36.8 & 97.9 & 72.9 & 74.7 & 68.7 & 83.0 \\
ColBERTv2 & \textbf{39.7} &	\textbf{98.4} &	\textbf{74.5} &	\textbf{82.6} &	\textbf{75.6} &	\textbf{84.3}\\
\Xhline{1pt}
\multicolumn{7}{l}{\emph{Generative Retrieval Models}} \\
\Xhline{0.3pt}
SEAL & 12.7 & — & — & — & — & — \\
MINDER & 18.6 & — & — & — & — & — \\
\Xhline{1pt}
\multicolumn{7}{l}{\emph{BM25 variants}} \\
\Xhline{0.3pt}
BM25 & 18.4 & 85.3 & 50.6 & 73.9 & 48.0 & 72.3 \\
 +RM3 & 15.7&	86.1	&52.2	&81.4	&49.0	&82.4\\
+Q2D & 21.4 & 91.8 & 66.2 & — & 62.9 & — \\
+HyDE & — & — & 61.3 & 88.0 & 57.9 & 84.4 \\
+LameR & — & — & 67.1 & \underline{\textbf{89.9}} & 62.7 & \underline{\textbf{88.7}} \\
+MUGI\sym{\tiny GPT3.5} & — & — & \underline{70.4} & — & 63.9 & — \\
+MUGI\sym{\tiny GPT4} & — & — & \underline{70.4} & — & \underline{64.4} & — \\
\Xhline{1pt}
\multicolumn{7}{l}{\emph{Ours}} \\
\Xhline{0.3pt}
Qwen3-base & 9.1 & 80.5 & 27.1 & 52.0 & 23.9 & 47.5 \\
\acronym{} & \underline{22.4}\sym{*}  & \underline{92.1}\sym{*}  & 63.1\sym{*}  & 82.1\sym{*}  & 60.8\sym{*}    & 81.5\sym{*}  \\
\Xhline{1pt}

\end{tabular}
\caption{In-domain evaluation on MS MARCO Dev, TREC DL’19, and TREC DL’20; \acronym{} is our main GRPO-trained model and Qwen3-base is directly prompting the LLMs for query specification generation. Best performing model is in \textbf{bold} and the best performing \emph{BM25 variant} model is \underline{underlined}. * = statistically significant difference between \acronym{} and BM25 with $p<0.05$}
\label{tab:indomain}
\end{table}

\begin{table*}[ht]
\centering
\scriptsize
\setlength{\tabcolsep}{5pt}
\begin{tabular}{lcccccccccc|cc}
\toprule
\textbf{Method} & \textbf{TC} & \textbf{NF} & \textbf{TO} & \textbf{DB} & \textbf{SF} & \textbf{SG} & \textbf{NW} & \textbf{RB} & \textbf{FQ} & \textbf{AG} & \textbf{Avg.} & \textbf{Our Avg}\\
\midrule
ANCE            & 65.4 & 23.7 & 24.0 & 28.1 & 50.7 & 24.9 & 38.2 & 39.2 & 30.0 & 41.5 & 36.8 & \textbf{47.1} \\

SPLADEv2 & 71.0	& 33.4&	27.2	&43.5	&69.3	&29.6	&39.4&	45.8	&33.6	&\textbf{47.9}&44.1 &\textbf{47.1}\\

ColBERTv2 & 73.8	& 33.8	& 26.3	& \textbf{44.6}& 	69.3& 	33.2	& 46.0& 	47.5	& \textbf{35.6}	&46.3	& 45.6&\textbf{47.1}\\
\midrule
\multicolumn{13}{l}{\textit{BM25 variants}}\\
BM25                    & 59.5 & 30.8 & 44.2 & 31.8 & 67.9 & 33.1 & 39.5 & 40.7 & 23.6 & 39.7 & 41.1 & \textbf{47.1} \\

\;\; + RM3 &59.3	&33.1&	35.0	&30.8	&64.6&	31.5	&42.6	&44.3	&19.2&	38.0	&39.8&\textbf{47.1}\\

\;\; + Q2D & 72.2 & 34.9 & 39.8 & 37.0 & 68.6 & —    & —    & —    & —    & —    & 50.5 & \textbf{53.1} \\
\;\; + HyDE & 69.1 & — & — & 36.8 & 69.1 & — & 44.0 & — & 27.3 & \underline{46.6} & 48.8 & \textbf{50.1}\\
\;\; + LameR & \underline{\textbf{75.8}}  & —    & —    & 39.0  & 73.5 & —    & 50.3 & —    & 25.8 & 40.2 &\textbf{50.8} & {50.1} \\
\;\; + MuGI\sym{\tiny GPT3.5}    & 69.7 & 36.0 & {46.3} & 41.2 & 72.0 & 35.8 & 46.6 & 49.7 & —    & —    & 49.6 & \textbf{50.2} \\
\;\; + MuGI\sym{\tiny GPT4}      & 72.9 & \underline{\textbf{37.4}} & 46.1 & \underline{42.7} &\underline{\textbf{74.0}} & \underline{\textbf{36.0}} & \underline{\textbf{50.0}} & 49.2 & —    & —    &\textbf{51.0} & 49.7 \\
\midrule
\textbf{\acronym{} (ours)} & 73.6\sym{*} 	&36.0\sym{*} &	\underline{\textbf{47.7}}&	38.8\sym{*} &	69.3&	34.8&	45.3\sym{*} &	\underline{\textbf{51.7}}\sym{*} &	\underline{27.5}\sym{*} 	&46.1\sym{*}  \\
\bottomrule
\end{tabular}
\caption{OoD performance (nDCG@10; higher is better) on selected BEIR datasets. A dash (—) indicates a value not reported. The rightmost columns show the macro-average across listed datasets for which a value is reported (Avg). We report the average of \acronym{} on the same subset of datasets (Our Avg). Best performing model is in \textbf{bold} and the best performing \emph{BM25 variant} model is \underline{underlined}. * = statistically significant difference between \acronym{} and BM25 with $p<0.05$}
\label{tab:statofart}
\end{table*}


\subsubsection{Efficiency Analysis}
\label{sec:efficiency}

\begin{figure}[t]
    \centering
    \includegraphics[width=1\linewidth]{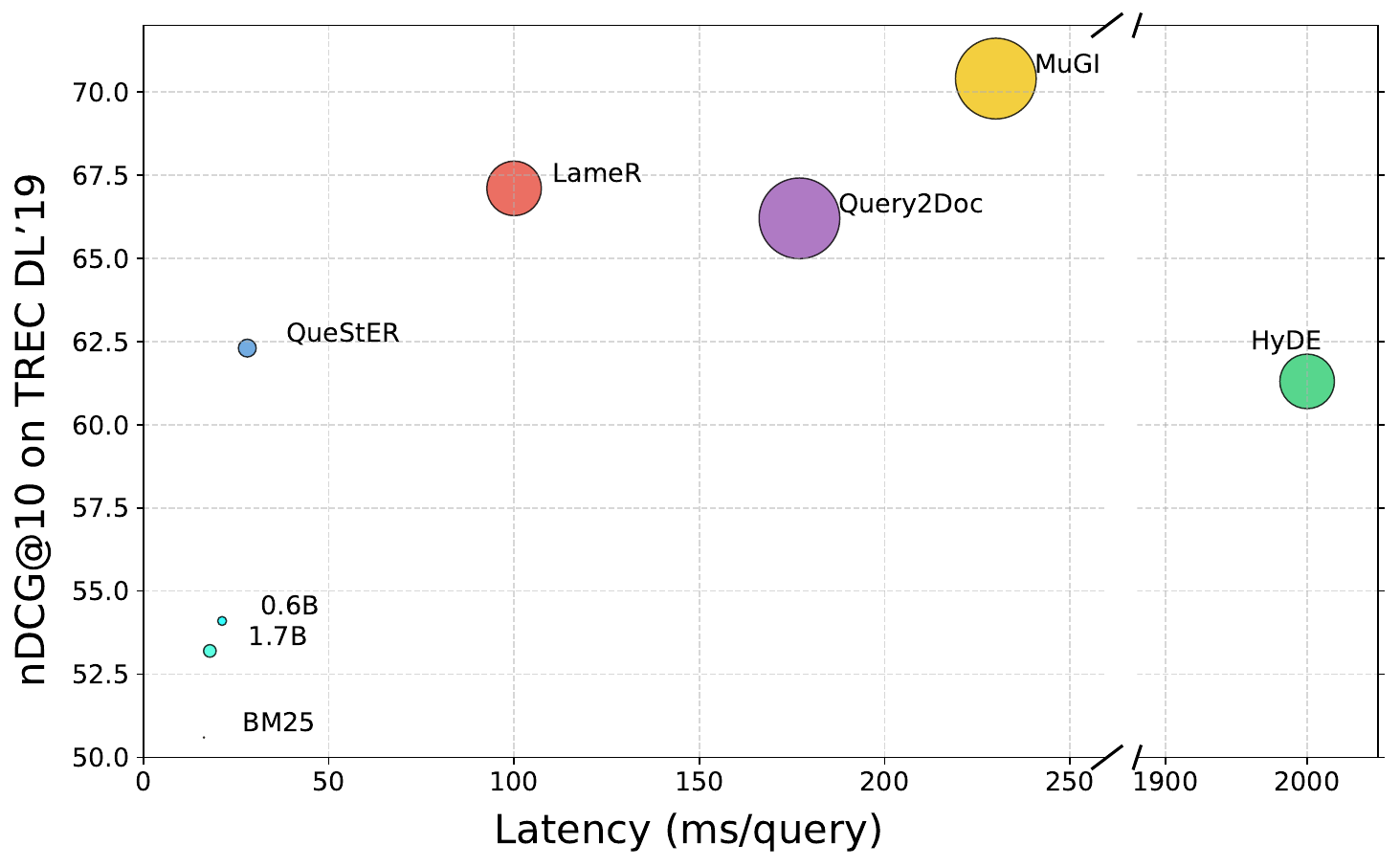}
 \caption{Trade-off between efficiency (ms/query, lower is better) and effectiveness (nDCG@10 on DL19, higher is better) for retrieval (generation time is not reported). 
Bubble size indicates model size (billions of parameters). 
Our \acronym{} offers a favorable balance, approaching MuGI and LameR in quality while being 4--7$\times$ faster.}
    \label{fig:qps_ndcg}
\end{figure}

Figure~\ref{fig:qps_ndcg} summarizes the efficiency-effectiveness trade-off of the different models~(using one thread when retrieving with Lucene,  on the 43 queries from DL'19). Leaving out the generation processing time, BM25 has the best efficiency (16.3 ms/query) but is less effective (nDCG@10 is 50.6). Our model \acronym{} has a strong balance (62.3 nDCG@10 with 28 ms/query). Our smaller variants lag in quality despite better efficiency. Prompt-expansion methods such as MuGI and LameR achieve a higher nDCG@10 but at the cost of efficiency (more than 100 ms/query).

\paragraph{Generation time}
\acronym{} generates a single keyword specification capped at 64 tokens using a small open model ($\leq4B$ parameters). In contrast, MuGI and LameR rely on large API-based LLMs (e.g., GPT-3.5/4, $\sim$170B) and can generate substantially longer expansions; for instance, MuGI may produce 4--5 rewrites of up to 512 tokens each per input query. As a result, end-to-end generation latency is not directly comparable across methods, and Figure~\ref{fig:qps_ndcg} reports retrieval-time latency only.





\subsubsection{Qualitative Analysis }

\begin{figure}[t]
    \centering
    \begin{subfigure}{\linewidth}
        \includegraphics[width=\linewidth]{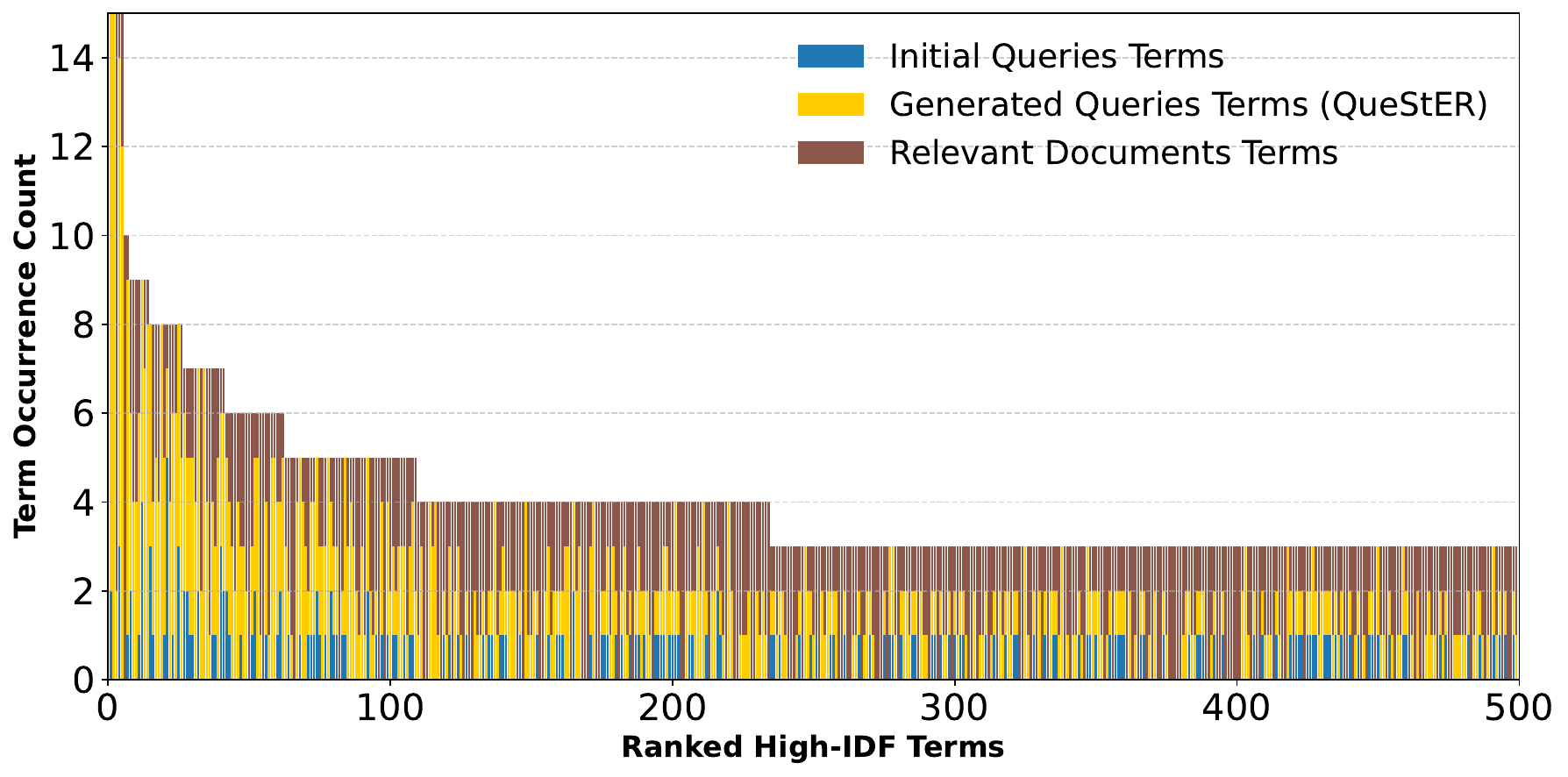}
        \caption{In-domain (TREC DL19 and DL20).}
        \label{fig:keyword_overlap_trec}
    \end{subfigure}
    \begin{subfigure}{\linewidth}
        \includegraphics[width=\linewidth]{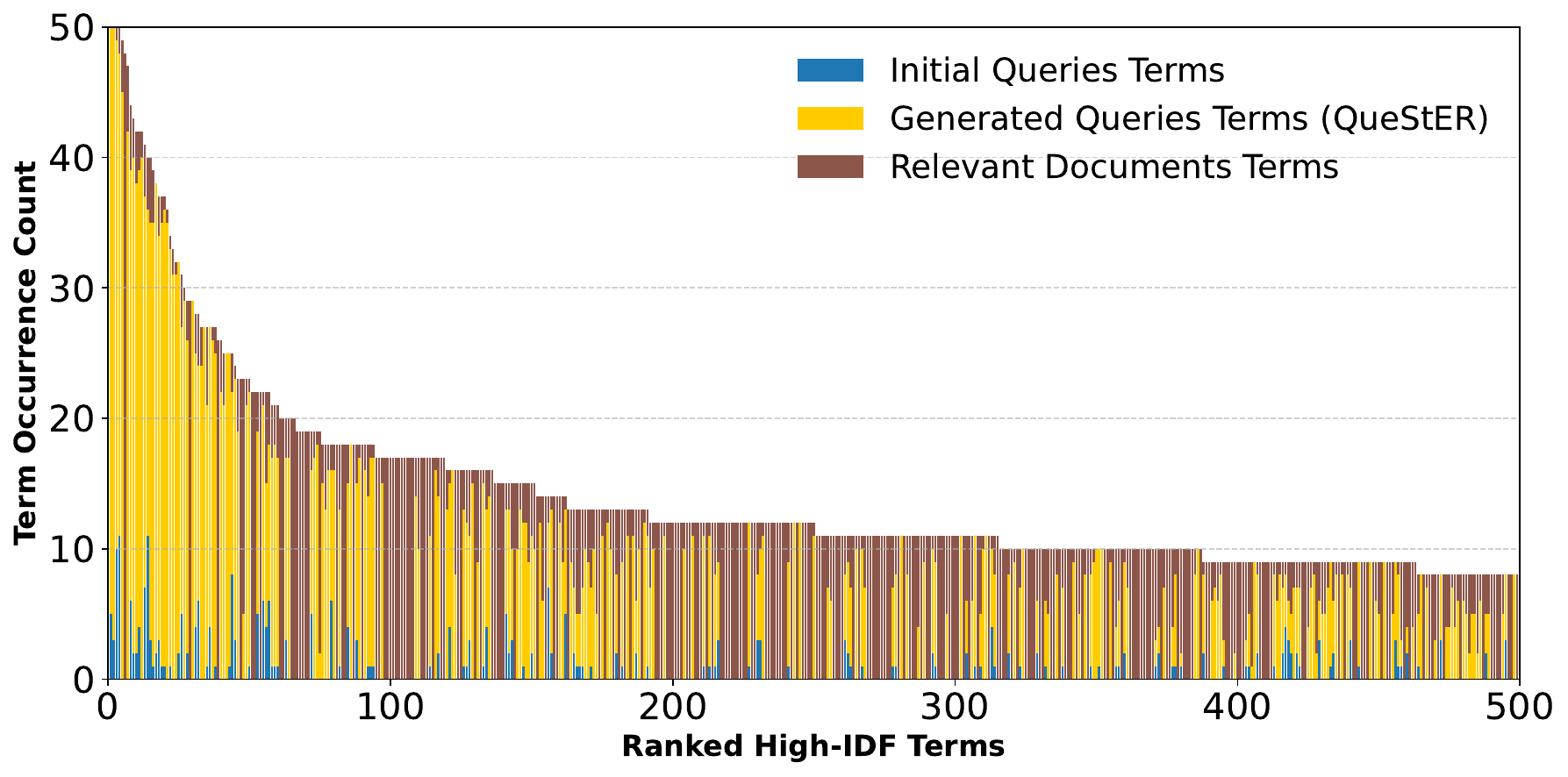}
        \caption{Out-of-domain (NFCorpus).}
        \label{fig:keyword_overlap_nfcorpus}
    \end{subfigure}
    \caption{Keyword overlap distribution between original queries (blue), generated queries (yellow), 
    and ground-truth relevant passages (brown). Generated queries consistently inject discriminative 
    vocabulary that aligns with relevant documents, both ID and OOD.}
    \label{fig:keyword_overlap}
\end{figure}

We analyze here the query reformulations that our model generates.
We analyze the overlap of high-IDF keywords in-domain (TREC DL'19/20) and out-of-domain (NFCorpus) in Figure~\ref{fig:keyword_overlap}. 
In both cases, the word frequency distributions align better between rewritten queries and relevant documents than with original queries.
On TREC DL, the overlap is especially pronounced, with generated queries closely mirroring the vocabulary of relevant documents. On NFCorpus, while the alignment is lesser due to domain shift, generated queries have a distribution that aligns much more with that of relevant documents. Together, these results reinforce our statement that \acronym{} query specifications not only enhance query–document matching in-domain, but also generalize effectively to unseen domains.
\\



An example of a query rewritten by \acronym{} on NFCorpus illustrates how the model expands a short, under-specified query into a richer set of semantically related phrases, improving lexical diversity and alignment with relevant documents:
\begin{description}
    \item[\textbf{Initial query:}] veggie chicken
    \item[\textbf{Generated query:}] veggie chicken recipe; chicken vegetable dish; veggie chicken healthy option; 
chicken stuffed with vegetables; veggie chicken recipe vegetarian; 
chicken vegetable mixture; veggie chicken salad; vegetable chicken recipe common; 
chicken vegetable used in vegetarian meal; veggie chicken substitute meat; 
chicken dish vegetable; vegetable chicken recipe.
    \item{\textbf{Final query (merging terms)}} chicken (×12), vegetable (×6), veggie (×5), recipe (×4), dish, salad, stuffed, healthy, mixture, substitute, meal.
\end{description}

\begin{table*}[t]
\centering
\scriptsize
\setlength{\tabcolsep}{5pt}
\begin{tabular}{llllllllllllllll|l}
\toprule
Model & \textbf{AG} & \textbf{CF} & \textbf{TC} & \textbf{DB} & \textbf{FV} & \textbf{FQ} & \textbf{HP} & \textbf{NW} & \textbf{NF} & \textbf{NQ} & \textbf{RB} & \textbf{SD} & \textbf{SF} & \textbf{SG} & \textbf{TO} & \textbf{Avg.} \\
\midrule
BM25 & 39.7 & 16.5 & 59.5 & 31.8 & 65.1 & 23.6 & 63.3 & 39.5 & 32.2 & 30.5 & 40.8 & 14.9 & 67.9 & 33.0 & 44.2 & 40.2\\
Qwen3-base & 29.6 & 17.9 & 65.1 & 24.9 & 40.5 & 21.6 & 33.6 & \textbf{47.9} & 31.3 & 31.0 & 35.5 & 12.2 & 67.2 & 16.3 & 34.1 & 33.9
\\
\rowcolor{black!10}
\textbf{\acronym{}} & \textbf{46.1}\sym{*} & {19.2}\sym{*} & 73.6\sym{*} & {38.8}\sym{*} & {70.3}\sym{*} & \textbf{27.5}\sym{*} & {64.3}\sym{*} & 45.3\sym{*} & 36.0\sym{*} & 43.0\sym{*} & \textbf{51.7}\sym{*} & 15.1 & 69.3 & {34.8} & 47.7 & {45.5}\sym{*}\\

\midrule
\multicolumn{17}{l}{\textit{(i) Effect of Model size }}\\
0.6B & 40.5 & 17.6 & 69.3 & 34.0 & 67.1 & 26.0 & 62.3 & 41.5 & 33.9 & 36.2 & 48.1 & 14.9 & 69.1 & 32.6 & 44.4 & 42.5\\
1.7B & 44.7 & 18.0 & 69.9 & 34.8 & 66.9 & 25.9 & 62.8 & 42.4 & 34.9 & 36.8 & 48.7 & \textbf{15.4} & 69.9 & 33.9 & 44.1 & 43.3\\


\midrule
\multicolumn{17}{l}{\textit{(ii) Supervised fine-tuning (SFT)}}\\
SFT & {33.9} & {16.7} & {59.5} & 25.2 & {60.8} & {20.5} & {60.4} & 38.7 & 30.7 & 30.9 & {42.1} & {14.3} & {70.3} & {27.4} & 33.6 & {37.2}\\
SFT+GRPO & 45.4 & 17.8 & 72.6 & 36.8 & 67.8 & 26.6 & 63.6 & 42.7 & 34.8 & 40.8 & 49.7 & 15.0 & 68.5 & 33.4 & 45.2 & 44.0\\

\midrule
\multicolumn{17}{l}{\textit{(iii) Hard-label assessments vs.\ CE gains}}\\

Hard & 45.7 & 17.8 & 71.8 & 38.3 & \textbf{70.7} & 27.0 & 63.3 & 44.8 & 35.6 & 42.7 & \textbf{51.7} & 15.3 & \textbf{70.9} & 34.3 & 46.9 & 45.1\\

Hard-w/o CE & 43.9 & 18.0 & 72.0 & 36.9 & 69.1 & 27.2 & 62.0 & 43.1 & 34.9 & 42.0 & 50.9 & 15.1 & 70.0 & 31.4 & \textbf{48.2} & 44.3\\
Soft-w/o CE & 44.5 & 17.1 & 72.0 & 37.1 & 67.5 & 26.8 & 62.8 & 44.4 & 35.3 & 41.3 & 48.9 & 15.1 & 70.3 & 33.4 & 46.4 & 44.2\\
\midrule
\multicolumn{17}{l}{\textit{(iv) Effect of standard deviation  $\nu$}}\\

$\nu = 1$ & 44.9 & 18.0 & 71.9 & 36.9 & 69.5 & 25.7 & 62.8 & 43.5 & 36.0 & 39.6 & 48.7 & 15.2 & 70.0 & 33.3 & 39.6 & 43.7\\
$\nu = 0.05$ & 44.8 &18.9 &72.2 &36.4& 68.5& 27.8& 62.7 &43.8 &33.1 &41.2 &48.9 &14.8 &70.3 &33.1 &41.9& 43.9\\
\midrule
\multicolumn{17}{l}{\textit{(v) KL weight $\beta$ ablation}}\\
KL5\% &38.6 &19.1 &64.4 &32.1 &65.8 &23.2 &58.9& 40.9 &33.6 &36.8 &46.1 &14.3& 68.8 &32.4 & 39.4 & 41.0\\
KL1\% &	41.1	&22.4&	66.7&	35.4&	70.5	&25.7	&62.1	&45.7&	35.5	&41.2&	52.4	&15.2&	71.2	&30.1&	37.9 & 43.5\\

\midrule
\multicolumn{17}{l}{\textit{(vi) Effect of Cut-off $k$ for nDCG@$k$}}\\

$k$=100 & 45.1 & 18.3 & 73.2 & 37.7 & 69.0 & 26.8 & 62.4 & 46.1 & \textbf{36.4} & 42.8 & \textbf{51.7} & 15.0 & 69.2 & 32.2 & 46.4 & 44.8\\

\midrule
\multicolumn{17}{l}{\textit{ (vii) concatenating original query during training vs.\ concatenating original query only in inference}}\\

Training & 44.9 & 18.5 & \textbf{74.5} & 37.5 & {70.3} & 27.4 & 62.7 & 44.5 & 36.2 & \textbf{43.6} & \textbf{51.7} & \textbf{15.4} & \textbf{70.9} & 33.0 & {47.8} & 45.3\\

Inference  & 44.3 & \textbf{19.4} & 74.0 & \textbf{39.1} & \textbf{70.4} & 27.3 & \textbf{65.5} & 45.1 & 35.9 & 42.5 & 51.1 & 15.2 & 69.9 & \textbf{35.8} & \textbf{47.9} & \textbf{45.6}\\

\bottomrule
\end{tabular}
\caption{nDCG@10 on 15 BEIR datasets with heterogeneous query profiles (datasets with many queries and longer descriptions vs.\ short keyword queries). The table reports per-dataset scores and macro-average; per-dataset maxima are in bold (ties bolded). \acronym{} is our main GRPO-trained model and Qwen3-base is directly prompting the LLMs for query specification generation. Ablations vary (i) model size (0.6B/1.7B/4B), (ii) the usage of SFT as a training starting point, (iii) supervision (hard qrels vs.\ CE-derived graded gains), (iv) SoftRank standard deviation $\nu$, (v) KL weight $\beta$, (vi) evaluation pool (BM25 top-100 only vs.\ top-100 plus CE-labeled documents within 10k) and (vii) concatenating the original query during training or inference. (statistically significant for compared to BM25 with $p=0.05$)}
\label{tab:results_NDCG10}
\end{table*}





\subsection{Ablations} \label{sec:Ablations}
In this section, we perform several ablation studies on various hyperparameters of our model. All the ablation results described below are reported Table~\ref{tab:results_NDCG10}. 

\paragraph{(i) Effect of Model size}

Since the number of parameters directly affects model latency (Figure~\ref{fig:qps_ndcg}), 
we examine the trade-off between efficiency and effectiveness across different parameter sizes (0.6B, 1.7B, and 4B). Performance improves consistently with scale. The 4B model achieves the highest average nDCG, followed closely by the 1.7B variant, while the 0.6B version lags behind. Although the 1.7B model offers a favorable latency–accuracy balance, the 4B configuration provides the most robust improvements.

\paragraph{(ii) Original/SFT Checkpoint Training Analysis}

RaFE~\citep{mao2024rafe} uses supervised fine-tuning (SFT) to warm up the model.
To investigate the effect of SFT in our case, we conduct experiments
in which we randomly sample 50,000 queries (from the MS MARCO training dataset, with no overlap with the queries used to train the model with RL). We use a large model (Qwen3-32B) to generate rewritten queries, generating 20 reformulated queries, keeping the best one \emph{if improving retrieval performance}. These couples (query, reformulated query) are used as a supervised dataset.

Using SFT improves over the backbone model, but does not reach BM25 performance. If we further train using GRPO (with SoftNDCG/CE), this further improves effectiveness but does not outperform the \acronym{} model without SFT.
Looking at generated queries, we hypothesize that using SFT narrows the rewrite space and reduces exploration.

\paragraph{(iii) Effect of different supervision source}

The results clearly show that replacing the sparse click-based labels with CE-derived labels produces more stable and effective learning.
Hard and Soft variants trained without CE perform worse (–1 nDCG@10), confirming that dense, graded CE supervision is crucial to guide GRPO toward consistent improvements as it is to train neural IR models.


\paragraph{(iv) Effect of $\nu$} \label{Variance}

The standard deviation $\nu$ in SoftRank (Section~\ref{softvshard}) controls the smoothness of pairwise document comparisons in SoftNDCG. As shown in Table~\ref{tab:results_NDCG10}, moderate smoothing with our main model \acronym{} that has $\nu{=}0.5$ achieves the best average nDCG@10 (45.5), outperforming both sharper $\nu{=}0.05$ and softer $\nu{=}1$ variants. A small $\nu$ makes the ranking signal too unstable, while a large $\nu$ over-smooths relevance differences, weakening the learning signal. These results suggest that moderate variance provides a stable yet discriminative reward for effective policy optimization.

\paragraph{(v) Effect of $\beta$}

We evaluated the impact of the KL-divergence weight (in GRPO) by testing $\beta\in\{0.0, 0.01, 0.05\}$. As shown in Table~\ref{tab:results_NDCG10}, setting $\beta=0.0$ (no KL regularization) yields the highest stability and overall nDCG@10 (45.5), while larger values consistently degrade performance (e.g., $40.9$  at $\beta=0.05$) . A higher $\beta$ overly constrains the policy to be close to the reference model, limiting exploration and reward optimization.

\paragraph{(vi) Effect of Cut-off $k$}

Finally, we need to set the cutoff value $k$ used to compute $\mathbb{E}(nDCG@k)$. We see that using $k=100$ consistently decreases the performance of the model, compared to the $k=10,000$ -- and this, even when assuming that documents not re-labeled by the CE are not relevant.


\paragraph{(vii) Effect of using the original query}

Following MuGI~\cite{zhang2024exploring}, we examine whether reusing the original user query helps retrieval. More precisely, we tried to add it after training (\emph{inference}), but this only provides a negligible gain (+0.1 in nDCG@10). Adding the query during both training and inference has a slight negative effect (-0.2 in nDCG@10). However, both strategies slightly improve the consistency of the performance across datasets, but more work is needed to understand exactly the effect of this modification.



\section{Conclusion and Future Work}

In this paper, we present \acronym, a generative retrieval approach that rewrites queries, which is trained with GRPO with a reward based on SoftRank~\cite{taylor2008softrank} and distillation. We show that \acronym{} has the best tradeoff between efficiency and effectiveness in its class of models (LLM-based query rewriting) and is competitive out-of the domain with neural IR models, even if it still lags behind when in-domain. This shows the potential of this type of approach -- we plan to explore more structured (and thus expressive) query languages as well as other optimization techniques in the future.




\section*{Limitations}
Public models (like Qwen or to a lesser extent cross-encoders) may contain data that overlap either with MS MARCO and/or BEIR. Measuring precisely how much data contamination is an issue we share with all the other works presented here.

In this work, we do not consider more structured query languages (e.g. with boolean, phrase, or proximity specifications) or hybrid dense–sparse backends. Although an interesting future avenue, this first work has shown that it was possible to improve existing related works in terms of efficiency while keeping almost the same effectiveness.

\section*{Acknowledgments}
The authors acknowledge Air Liquide and  ANR - FRANCE (French National Research Agency) for its financial support of the GUIDANCE project n°ANR-23-IAS1-0003. 
This work was granted access to the HPC resources of IDRIS \& CINES (Project No. A19) under the allocation made by GENCI.




\bibliography{custom}

\appendix


\end{document}